\documentclass[onecolumn
reprint, 
 amsmath,amssymb,
 aps, physrev,
floatfix,
]{revtex4-2}
\usepackage{lipsum, babel}

\usepackage{graphicx}
\usepackage{dcolumn}
\usepackage{bm}
\usepackage{mathrsfs}
\usepackage{placeins}
\usepackage{rsfso}
\usepackage{hyperref}
\usepackage{amsmath}
\usepackage{float}
\usepackage{subcaption}

\begin{document}

\preprint{APS/123-QED}
\title{\textbf{Spatio -- Temporal Weak Measurement of Chiral Ultra short Laser Pulse}} 

\author{
Sahil Sahoo$^{1}$,
Andre Yaroshevsky$^{1}$,
Dima Cheskis$^{2}$,
Yuri Gorodetski$^{1,3,*}$
}

\affiliation{$^{1}$Department of Electrical and Electronic Engineering, Ariel University, Ariel 40700, Israel}
\affiliation{$^{2}$Department of Physics, Ariel University, Ariel 40700, Israel}
\affiliation{$^{3}$Department of Mechatronics and Mechanical Engineering, Ariel University, Ariel 40700, Israel}

\date{\today}
             \email{Contact author: yurig@ariel.ac.il}

\begin{abstract}
We present a comprehensive study on the spatio-temporal weak--measurement of a chiral ultrafast optical pulse. We create a chiral vector wave packet by transmitting ultrashort laser pulse via a birefringent or magneto-optic medium. Employing time-resolved leakage radiation microscopy, we examine how the real and imaginary components of the weak--value parameter (\( \epsilon \)) influence pulse propagation over time. Our technique allows us to detect and categorize the temporal polarization fluctuation in a $75 fs$ pulse with an excellent repeatability. The achieved experimental results demonstrate a satisfactory consistency with the theoretical predictions.   
\end{abstract}
\maketitle
\section{Introduction}
A transverse shift of a spatially confined optical beam, which depends on its polarization, is known as the Spin Hall Effect of Light (SHEL) \cite{PhysRevD.5.787,onoda2004hall}. This phenomenon arises from the spin-orbit interaction (SOI) of light—an interaction between the intrinsic and extrinsic momenta of photons \cite{bliokh2015spin,kim2023spin}. As a result, photons with opposite spin states (i.e., handedness of circular polarization) can follow different trajectories \cite{o2014spin}. The SHEL has been experimentally observed during refraction and reflection at planar optical interfaces, manifesting as a polarization dependent transverse shift of the beam centroid perpendicular to the plane of incidence. It can also occur due to refractive index gradients, which act analogously to electric potential gradients, causing the beam to split into its circular polarization components upon refraction \cite{onoda2004hall,Ling_2017,PhysRevLett.123.243904}.
Furthermore, this effect has been demonstrated using surface plasmons (SPs), where a plasmonic slit can detect polarization-dependent shifts in the transmitted beam \cite{PhysRevLett.109.013901, zia2007surface}.
Although the magnitude of the SHEL is typically on the order of a small fraction of the wavelength, weak measurement (WM) techniques can be applied to a photon spin (helicity) \cite{strubi2013measuring,PhysRevLett.60.1351,PhysRevD.40.2112,PhysRevB.76.155324,PhysRevLett.94.220405,PhysRevResearch.5.023048,PhysRevLett.74.2405,PhysRevA.84.043806,rebufello2021anomalous} to amplify this small shift to observable levels \cite{hosten2008observation}. This technique has been proven to be a powerful tool for high-precision measurements and optical interface studies, with precision, sometimes reaching the angstrom scale \cite{kim2022reaching,zhang2015precision,magana2014amplification,wang2020probing,lee2025real}. Furthermore, measurement theory accurately characterizes frequency and time shifts in the near-zero scattering regime, offering a unified framework applicable to both classical and quantum regimes \cite{asano2016anomalous,pan2023weak}. 

The classical interpretation of the weak measurement describes the interaction between a transversely confined polarized optical beam and a planar interface by the SOI hamiltonian, probing the helicity of the incident light. By strategically placing a linear polarizer before the interface, the incident state is pre-selected with respect to the plane of incidence. The SOI, resulted from the polarization dependent Fresnel coefficients at the interface refraction produces a weak transverse displacement originally known as the Imbert shift \cite{PhysRevD.5.787}. The post-selection made by an additional polarizer, almost perpendicular to the first one, results in a significant (\(\approx 20\)-fold) amplification of this spin-dependent beam shift \cite{YuWangLiZhaoZhouQuSong+2021+3031+3048,Qiu:16,rodriguez2010optical, PhysRevLett.94.220405,ChoiShimKimTangLiRhoLeeKim+2024+3877+3882}. In fact, the role of the pre-selection is to create a balanced mixture of two eigen states. The weak interaction leads to a tiny polarization dependent shift between them while the post-selection almost fully removes the mixed state giving rise to the measurement enhancement.
Almost a decade ago, it has been demonstrated that a coupling of a tightly focused optical beam to SPs by a single slit naturally implements the weak measurement scheme, where the fixed linear polarization of the excited SP mode serves as an intrinsic post-selection mechanism \cite{PhysRevLett.92.107401}. By using slightly tilted linear or slightly elliptical input polarizations, both real or imaginary components of the weak value of spin could be extracted via the measurement of spatial or angular transverse shifts respectively \cite{PhysRevA.80.061801, bliokh2013goos,qin2009measurement}.

In the early 90s a concept of a \textit{temporal} WM has been proposed \cite{PhysRevLett.64.2965} where a weak polarization-dependent shifting of eigen-state time-evolutions could be significantly enhanced by an appropriate post-selection. This effect has been introduced as a quantum time-translating machine. In optics, a femto-second polarized laser pulse can be considered as the superposition of the eigen-state time-evolutions and the weak measurement operator can be applied by a birefringent medium. 
In such a material the initial pulse is separated in time into the ordinary and the extraordinary modes due to a group velocity difference.
The emerging superimposed eigen-state time-evolutions effectively represent a chiral pulse with a weakly varying (rotating) polarization.   
Similarly to the spatial WM a specific post-selection operator
can be applied to perform as the strong measurement on a time axis. However, such a technique requires a very high temporal resolution and a signal-to-noise ratio of the detector \cite{gorodetski2016tracking}. 
\begin{figure*}[ht]
\centering
\includegraphics[width=1\textwidth]{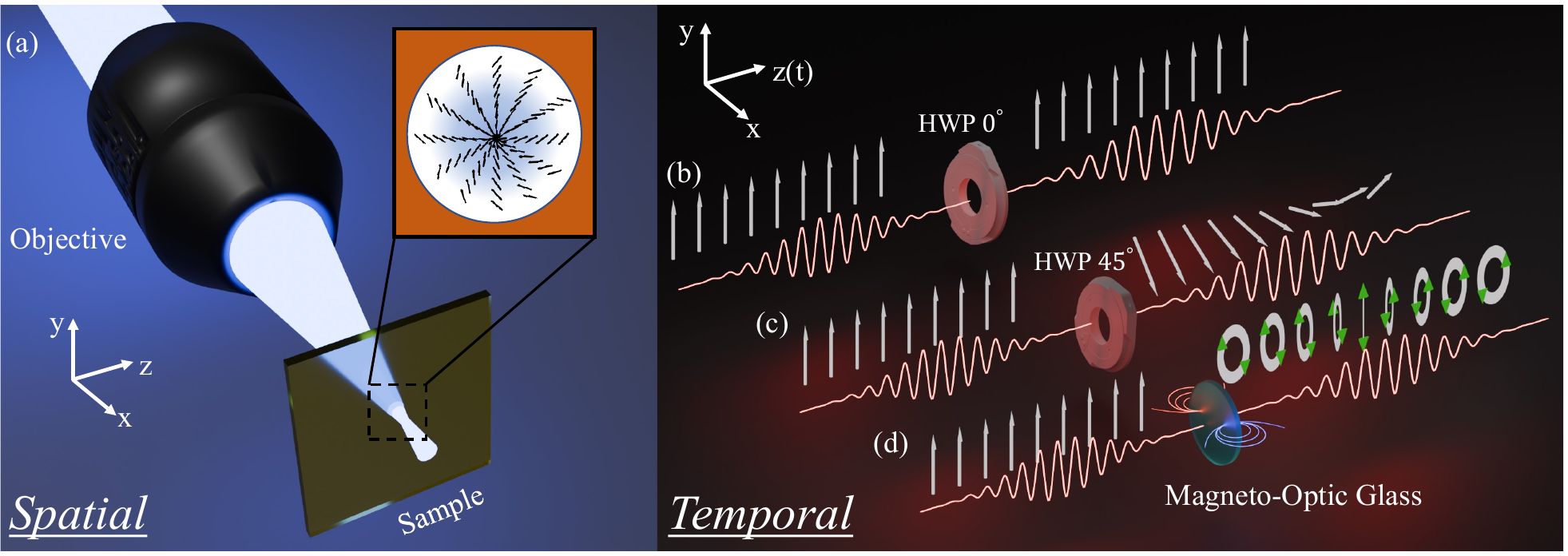}
\caption{\label{figure1} (a) Polarization variation in space due to a tight focusing. (b),(c) Temporal polarization modulation for a pulse passing through a HWP at  $0^\circ$ and at $ 45^\circ$. (d) Polarization state variation due to a Faraday effect.}
\end{figure*}

In this letter we demonstrate a combined effect of the spatial and temporal confinement resulting in a ``spatio-temporal weak measurement''. A femtosecond linearly polarized laser pulse is transmitted through a birefringent medium where temporal splitting into ordinary and extraordinary eigen-states occurs. 
The emerging mixed-state time-evolution is then tightly focused by a high-$NA$ (Numerical Aperture) objective (see Fig. \ref{figure1}a) on a plasmonic slit which provokes a spin-orbit interaction in analogy to our earlier work \cite{PhysRevLett.109.013901}.
However, the time varying polarization state caused by the birefringence can be regarded as a time-dependent weak value. As can be seen from Fig. \ref{figure1}b depending on the type of the medium birefringence (linear or circular) the polarization state changes its orientation (real weak-value) or ellipticity (imaginary weak-value) inducing the pulse chirality.
The slit then plays a role of the \textit{spatial} post-selection while the time evolution of the states is tracked by our time-resolved leakage radiation microscopy (TRLRM) system, presented in Fig. \ref{figure2}a.
To obtain the real weak-values we placed a HWP at $\pm 45^\circ$ with respect to the incident polarization and for the imaginary weak-values a Faraday medium with an external magnetic field applied along or against the propagation axis was used. 
Under this configuration WM-induced angular or spatial shifts of the SP pulse are clearly observed. Tracking of the plasmon trajectories in time enables to measure the weak value across the whole complex plane for the time-shifts of the order of $1\%$ of the pulse duration. 

Finally, we employ a chiral liquid crystal plate as the WM operator in our system~\cite{goodby1991chirality}. Despite the apparent symmetry, two distinct outcomes are detected, revealing the intrinsic optical activity imparted to the pulse. Notably, simultaneous observation of angular shifts and spatial displacements suggests that under specific retardation conditions the liquid crystal exhibits both linear and circular birefringence. These findings highlight the potential of time-resolved polarization dynamics for applications in optical sensing and telecommunications \cite{vanwiggeren2002communication, shevchenko2017polarization}, as well as for real-time detection of circularly or linearly polarized light \cite{li2015circularly, ran2021integrated, khairulin2022amplification}.
\section{\label{Experiments}Spatio-temporal weak measurement}
\begin{figure*}[ht!]
\centering
\includegraphics[width=130mm]{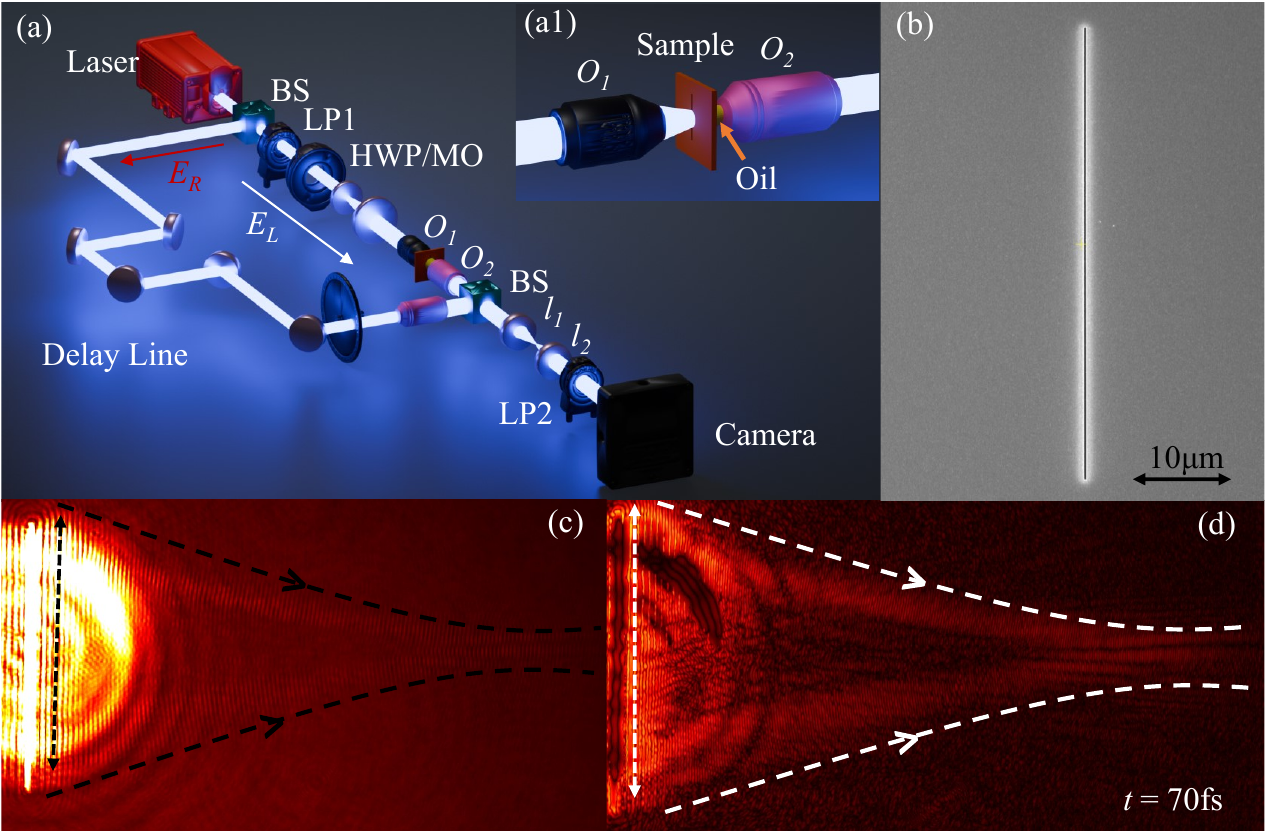}
\caption{\label{figure2} Experimental Setup and the results. (a) Schematic diagram of the experimental setup with BS - beam splitter, LP - linear polarizer, $O_1$ and $O_2$ - objectives and $L_1$ and $L_2$ lenses.  (a1) Leakage radiation microscopy setup. (b) SEM image of the slit. (c) Leakage radiation (LR) image recorded at $t = 70$ fs with a perfect TE polarization. (d) Processed pulse image after filtering.}
\end{figure*}
\begin{figure*}[ht!]
\centering
\includegraphics[width=80mm]{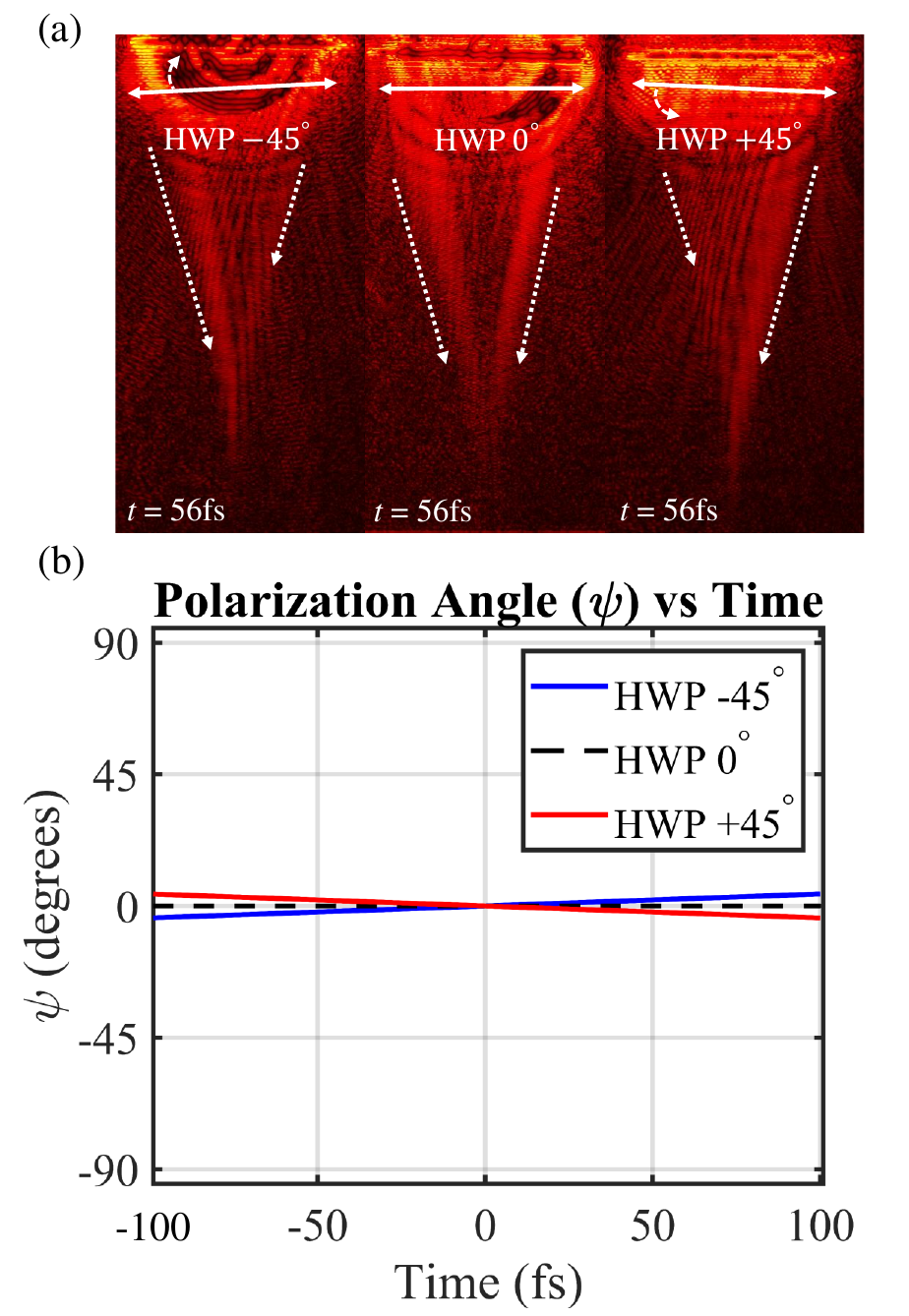}
\caption{\label{figure3} Plasmonic distribution after the post-selection. (a) Measured time-snapshot of the plasmonic beams with the HWP rotated at $45^\circ$, $0^\circ$ and $-45^\circ$. (b) Polarization rotation angle variation in time calculated using Eq. \ref{ShortEb} for the corresponding cases of $\epsilon$.}
\end{figure*}

The scheme of our setup is illustrated in Fig. \ref{figure2}a. A femtosecond pulsed laser (MENLO C-fiber 780/\(\lambda_0 = 785\,\mathrm{nm}\)) average power 100mW ) with a pulse width of approximately $75fs$ is employed. A beam-splitter (BS) is used to separate the incident light into the reference ($E_R$) and the leakage ($E_L$) paths. The $E_L$ signal passes through a vertical linear polarizer to ensure a pure pre-selected state, and then a weak-measurement is performed by a HWP at $\pm 45^\circ$ or a magneto-optic material (MO). The emerging pulse is collimated and focused onto a single $200 nm$ wide slit etched within the $80\,\mathrm{nm}$ thick gold layer evaporated on a $160\,\mu\,\mathrm{m}$ thick glass cover slip (the scanning electron microscope-SEM image of the slit is shown in Fig. \ref{figure2}b). The leakage radiation is extracted from the back side of the sample by use of an oil-immersion objective $O_2$  (wih $NA$ = 1.25/ $\times$100 magnification) attached to it with an index-matching oil (see inset a1 in Fig. \ref{figure2}a). The lens assembly $l_1$, $l_2$ in combination with the imaging infinity-corrected objective $O_2$ create an real-space image of the plasmonic signal excited by the slit in the camera. The reference signal $E_R$ passes via a delay line by which an optical path difference can be varied. 
This beam is phase-corrected attenuated and then recombined with the leakage signal. The interference image is captured by the camera.
Additional horizontal linear polarizer is used to filter out scattering contributions.        
A typical interference image is presented in Fig. \ref{figure2}c. We use a post-acquisition image processing described in \cite{gorodetski2016tracking} to isolate a temporal plasmonic signal (shown in Fig. \ref{figure2}d) for each time delay in order to track the plasmonic pulse propagation dynamics.
When comparing the results obtained in similar times for the HWP at $\pm 45^\circ$ we notice a strong assymetry (see Fig. \ref{figure3}a). Moreover when we investigate an image of the plasmonic distribution obtained with the HWP at $0^\circ$ (we have actually measured this image with $\left| Y\right>$ polarization and a vertical slit) the beam exhibits a full symmetry and a well-pronounced phase dislocation line in the center. In the following discussion we associate these clear differences with the weak measurement of the polarization state variation in time induced by the retarder.  
The initial laser pulse time dependence can be expressed as, 
\begin{equation}
E(t) = \frac{E_0}{2} e^{-2\ln 2\frac{(t-\tau)^2}{T^2} } e^{i \omega_0 (t-\tau)} \left| Y \right>,
\label{Gauss}\end{equation} 
where $E_0$ is the peak amplitude, $\omega_0$ is the central frequency, $T$ is the time duration of the pulse and the initial polarization state is represented in the linear basis, where $\left| Y \right> = [0,1]^T$ and $\left| X \right> = [1,0]^T$.
The propagation of the pulse in space is registered by the $\tau = z/c$ with $z$ being a position of the pulse center and $c$ the velocity of light. 
We assume that the birefringent material of the HWP rotated at $\pm 45^\circ$ equally decelerates the group and the phase velocity of the pulses leading to emerging signal of the following form,
\begin{equation}
\left| E_{B} \right> = \frac{1}{\sqrt{2}}\left | \nearrow \right>E(t-\frac{1}{4}t_c) + \frac{1}{\sqrt{2}}\left | \searrow \right>E(t+\frac{1}{4}t_c)
\label{Combined}\end{equation}
where the ordinary and extraordinary polarization states are represented by $\left | \nearrow \right> = \left [1,1\right ]^T$ and $\left | \searrow \right> = \left [1,-1\right ]^T$, respectively and $t_c = \frac{2\pi}{\omega_0}$ stands for the wave period.
Substitution of the corresponding time-shifts into Eq. \ref{Gauss} leads to the $\pi$-phase delay in the oscillating part of the field (as expected from the HWP) so it can be regarded as a rotation of the reference frame by $90^\circ$. Accordingly the resulting combined field after the HWP can be evaluated by placing a ``$-$'' sign between the terms in Eq. \ref{Combined}.
This calculation yields  
\begin{equation}
\left| E_{B}(t) \right> \propto \mathscr{A}(t)e^{i\omega_0 (t-\tau)} \left [\begin{matrix}
\cosh \left( \frac{\ln 2 (t-\tau) t_c}{T^2} \right) \\
\sinh \left( \frac{\ln 2 (t-\tau) t_c}{T^2} \right)
\end{matrix}\right ].
\label{Eb}\end{equation}
The ``modified'' Gaussian envelope of the pulse,  $\mathscr{A}(t) = e^{\frac{-2\ln 2}{T^2}((t-\tau)^2+\frac{ t_c^2}{16})}$ shows a slight widening, that will be neglected hereafter in the sake of simplicity. By considering that $t_c<<T$ and absorbing the harmonic part in $\mathscr{B}(t)$ Eq. \ref{Eb} can be reduced to   
\begin{equation}
\left| E_{B}(t) \right> \propto \mathscr{B}(t) \left( \left| X\right> - \epsilon (t) \left| Y \right>\right),
\label{ShortEb}
\end{equation}
where the small number, $\epsilon = \frac{ln 2 (t-\tau) t_c}{T^2}$ represents a linear time-dependent polarization variation.
The polarization angle  $\psi(t) = \tan^{-1} \epsilon(t)$ is calculated \cite{Comm} for the time range of 200 fs for three proposed orientations of the HWP (see Fig. \ref{figure3}b.). By approximating $\psi(t)$ using Taylor expansion to the first order we find a very weak linear slope of $3.2 \times 10^{-4} [fs^{-1}$] as can be observed in the figure. The comparison of the pulse peak amplitude with the residual $y$ polarization component is explicitly described in Appendix A.

Representing this expression in the frequency domain yields,
\begin{equation}    
\left| \tilde{E}_{B}(\omega) \right> \propto \mathscr{C}(\omega) \left( \left| X\right> - i \tilde{\epsilon} (\omega) \left| Y \right>\right),
\end{equation}
where $\mathscr{C}(\omega) \propto \exp\!\left[ \frac{T^2(\omega-\omega_0)^2}{2 }\right]$ is the Gaussian wavepacket envelope in the frequency domain, and $\tilde{\epsilon}(\omega) = \ln 2t_c(\omega-\omega_0)$. 
Interestingly, the problem now reduces to the summation of the plasmonic wavefronts excited by different harmonics with corresponding frequency dependent ellipticity. We suppose that each of the harmonics has a spatial Gaussian profile, $\Phi (x) \propto [-x^2/\delta^2]$, with the waist of $\delta \sim (k_0 NA)^{-1}$ defining the ``pointer'' width (here $k_0 = \omega_0/c$ is the incident light mean wavenumber and $NA =0.3$ is the numerical aperture of the focusing objective $O_1$).
\begin{figure*}[h!]
\centering
\includegraphics[width=1\columnwidth]{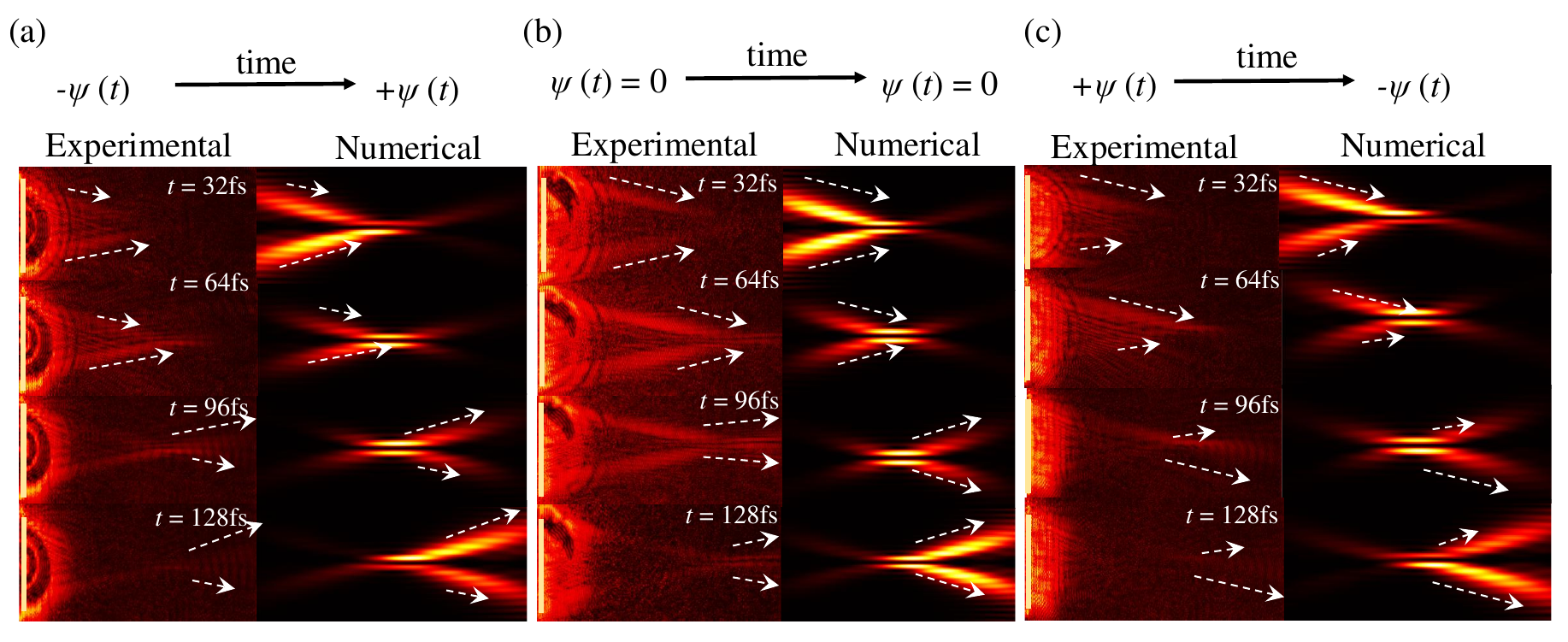}
\caption{\label{figure4} Experimental and numerical results showing the angular shift of the pulse over the time interval from \( t = 32\,\text{fs} \) to \( 128\,\text{fs} \). The results represent measurements done with (a) HWP at $-45^\circ$, (b) HWP at $0^\circ$ and (c) HWP at $+45^\circ$. White arrows guide the eye to the angular asymmetry of the intensities.}
\end{figure*}

As has been shown in \cite{PhysRevLett.109.013901} the spatial weak-measurement achieved by the tight focusing followed by post-selection performed by a plasmonic slit results in a momentum shift of $\Delta k \propto (k \tilde{\epsilon}\delta^2)^{-1}$. While this is the case for a single frequency component, the combined contribution from the whole spectrum range of the pulse appears to depend upon its bandwidth. We used a straight forward plane wave expansion method with time-varying ellipticity to model the tight focusing followed by a polarization post-selection imposed by the slit \cite{PhysRevLett.109.013901}. 
The results of the numerical simulation based on this method and taking a pulse duration time of  $75\,\mathrm{fs}$ are presented in Fig. \ref{figure4} for four time instants along with the results measured by the TRLRM system. The snapshots of the pulse propagation were done in intervals of $32\,\mathrm{fs}$ for three HWP orientations ($+45^\circ$, $0^\circ$ and $-45^\circ$). The beams are excited by nearly TE polarization, so the plasmonic field vanishes around the optical axis. The polarization angle variation in time is schematically depicted in the top part of the images. Noteworthy, the calculated and the measured distributions mainly exhibit an angular displacement where the intensity shifts from upper (lower) part to the lower (upper) branch. In addition we observe some spatial shifting in the focal region. Interestingly, when we increase the pulse duration time, the calculated spatial shift becomes dominant as expected from the weak value measured with the CW laser illumination (refer to appendix A). 

In the next stage we replaced the HWP by MR3-2 magneto-optic glass and used an external magnetic field of $H = 300 \text{mT}$ to induce a Faraday effect (refer to Appendix B). The magnetic field applied parallel or antiparallel to the optical axis generates a circular birefringence resulting in the relative retardation of the circular polarization states, $\left| R \right> = \frac{1}{\sqrt{2}}[1, -i]^T$ and $\left| L \right> = \frac{1}{\sqrt{2}}[1, i]^T$ \cite{born2013principles}. 

Accordingly, one can expect the appearance of a tiny ellipticity, increasing towards the pulse edges and a somewhat rotated linear polarization around the peak amplitude.
\begin{figure*}[ht!]
\centering
\includegraphics[width=75mm]{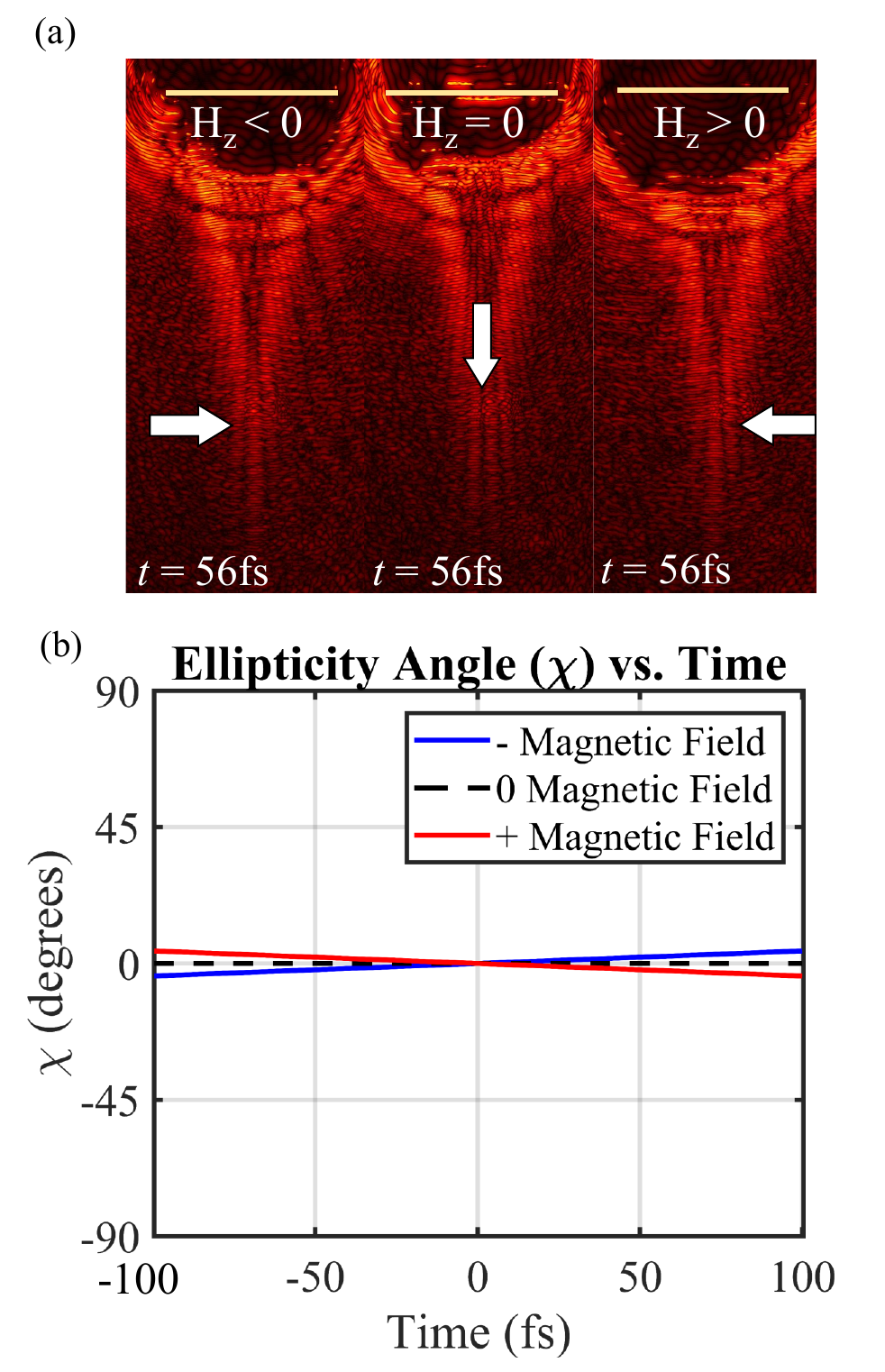}
\caption{ Plasmonic distribution after the post-selection of a pulse transmitted through a Faraday medium. (a) Measured time-snapshot of the plasmonic beams with a parallel, antiparallel or null magnetic field. (b) Calculated polarization ellipticity variation in time for the corresponding cases of $\epsilon$. 
}
\label{figure5}
\end{figure*}
We can, therefore, use our approach after (a) -- assuming that the slit is pre-aligned to be parallel to the orientation of the linear polarization at the pulse peak and (b) -- substituting $i\tilde{\epsilon} (t)$ instead of $\epsilon(t)$ into Eq. \ref{ShortEb}. This way, the signal is characterized by a time-dependent ellipticity, $\chi(t) = \tan^{-1} \tilde{\epsilon(t)}$. 
\begin{figure*}[ht!]
\centering
\includegraphics[width=1\columnwidth]{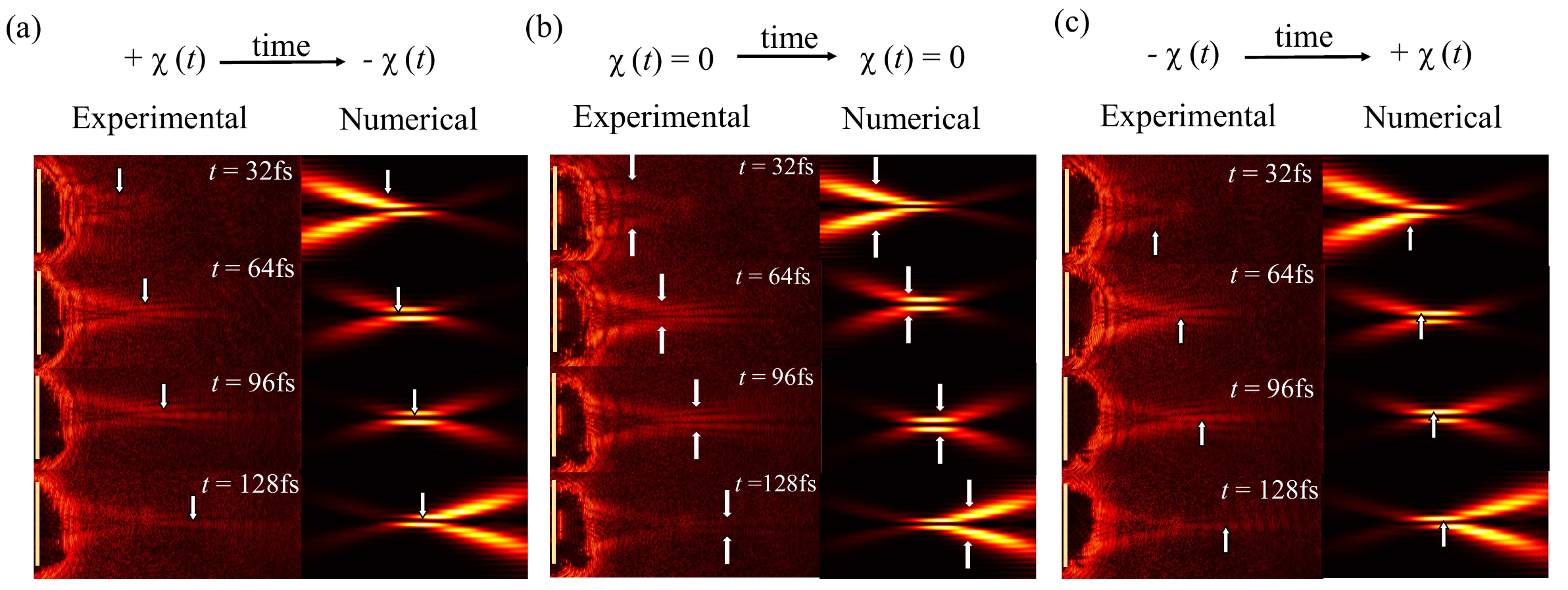}
\caption{Experimental and numerical results showing the spatial shift of the pulse over the time interval from t = 32 fs to 128 fs. The measurements represent the cases of (a) \( H_z>0\), (b) \( H_z=0\) and (c) \( H_z<0\). The white arrow points to the shift direction.}
\label{figure6}
\end{figure*}
The distribution of the plasmonic pulse behind the slit measured with the parallel ($H_z>0$), zero ($H_z=0$) and anti-parallel ($H_z<0$) magnetic field is presented in Fig. \ref{figure5}a. All the measurements have been done at the same relative time ( $56 fs$) of the pulse propagation and with a properly calibrated pre-selected state. We notice that the pulse with the applied magnetic field is mainly characterized by a shifted focal spot (marked by a white arrow). The calculated ellipticity value is shown in Fig. \ref{figure5}b for the cases under study. We note, that in contrast with the classical Faraday effect where only the rotation of the polarization, $\psi$ can be measured by the Malus' law, our approach performs an ultra-sensitive measurement of the ellipticity $\chi$. Moreover, as before, when the polarization evolution is reversed by flipping the direction of the magnetic field, the focal spot shifts in the opposite direction (see Fig.\ref{figure6}). We use the TRLRM scheme to achieve plasmonic pulse propagation in time in order to compare the three discussed configurations of the magnetic field with the calculated model for the time-dependent ellipticity. Notably, while the focal spot shifting is clearly observable, the angular distribution of the plasmonic pulse stays almost symmetric. The explicit measurement scheme is described in more detail in Appendix B.
The general time-dependent polarization state can be represented by a path on a Poincar\'{e} sphere running along a meridian or a latitude line, respectively (see Fig. \ref{poincare}).
\begin{figure*}[ht!]
\centering
\includegraphics[width=120mm]{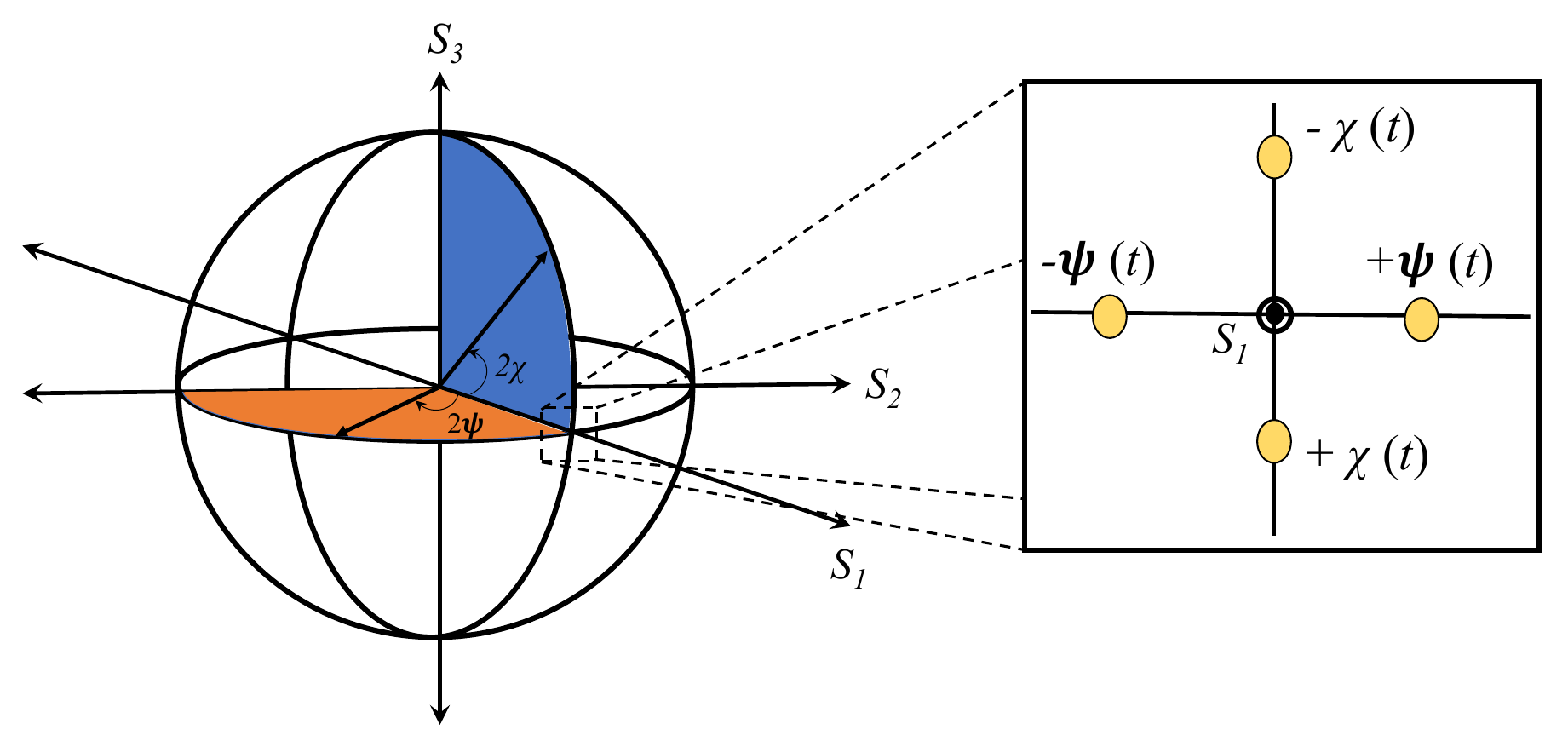}
\caption{ Polarization variation path represented on a Poincar\'{e} sphere. Circular/linear birefringence is responsible for the change in ellipticity/polarization rotation angle in time.}
\label{poincare}
\end{figure*}
Apparently, the WM scheme shows a capability of independently measuring these time variations in the pulse state.
To demonstrate the combined effect of the $\psi (t)$ and the $\chi(t)$ we use a $Q$-plate  (a liquid crystal phase plate) with externally variable retardation (see details in appendix B). 
The chiral nature of the liquid crystal under applied voltage results in a general optical activity simultaneously modifying $\psi$ and $\chi$ whose values are interrelated through the Kramers-Kronig relations \cite{Engheta_KK}. 
The ultrashort laser pulse passing through it experiences a combined polarization variation, that can be described by setting $\epsilon(t)$ as a general complex number.
The plate was biased at $1.3 V$ to induce elliptical polarization and the orientation of the polarization ellipse was accurately tuned to correspond to the specific post-selection imposed by the slit. 
\begin{figure*}[ht!]
\centering
\includegraphics[width= 0.5\columnwidth]{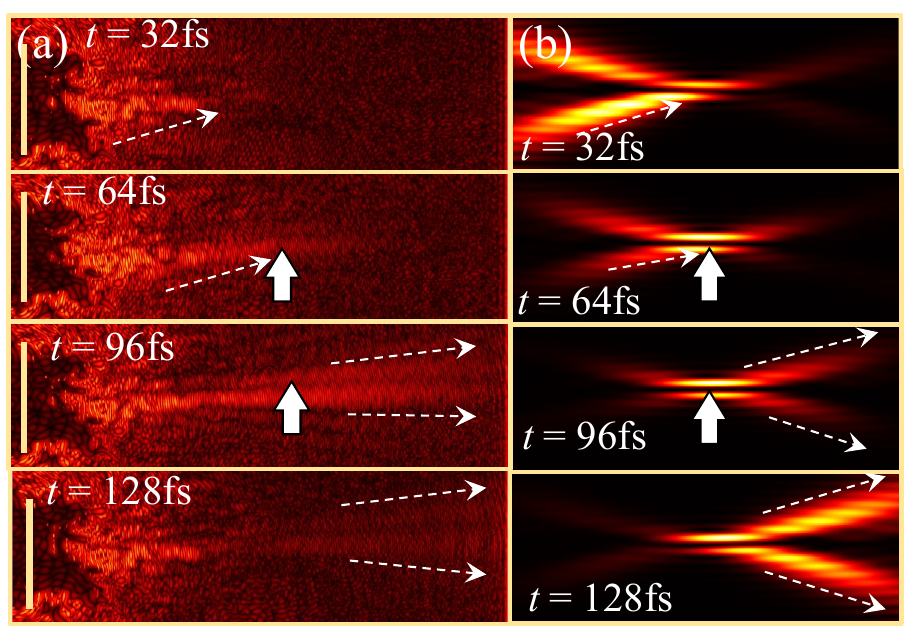}
\caption{ (a) Measured and (b) simulated results of the spatio-temporal WM with a combined effect of $\psi$ and $\chi$ varying in time. Experimental results showing the temporal evolution of the pulse from \( t = 32 \, \text{fs} \) to \( t = 128 \, \text{fs} \) under the polarization transformation \( -\chi, -\psi \rightarrow +\chi, +\psi \) resulted from a $Q$-plate biased with $1.3$ V. The calculation was performed by a proper fitting of a complex value of $\epsilon(t)$. The white arrows emphasize the beam angular and spatial shifts.}
\label{figure8}
\end{figure*}
Consequently, experimental observation of pulse propagation shows a simultaneous angular and spatial shifting of the beam, as shown in Fig. \ref{figure8}. We also perform a numerical calculation using our proposed method while using a complex-valued $\tilde{\epsilon}(t)$ in Eq. \ref{ShortEb}. And the experimental data shows a good correspondence to the calculated results. It is noteworthy that the measured snapshots reveal an interesting dynamics of the propagating plasmonic pulse. In particular, it is possible to see how the symmetry varies in time due to the temporal weak-value dependence. This is the main advantage of the TLRM system in the application of the spatio-temporal WM. 

\section{\label{Conclusion}Conclusions}
In summary, we have experimentally demonstrated the spatio-temporal WM of  ultra-fast optical pulses, with a weak temporal evolution of the polarization state. We employed a TRLRM technique to obtain time-resolved plasmonic distribution. The weak measurement post selection was done by a narrow slit efficiently performing as a polarizer. Our analysis revealed that due to the limited bandwidth of the incident light pulse the temporal WM applied by a birefringent material can be revealed by our simple technique with an impressive precision. We have demonstrated the effects of the linear and the circular birefringence and also their combination. We believe, that beyond the physical significance, our method can be implemented for ultrasensitive measurements of polarization states weakly varying in time. This opens a door for novel nanophotonic devices that may find valuable applications in biomedical optics, sensing, optics communications and many more. 
\newpage
\appendix
\section{Retardation in a Pulse vs. a Continuous wave}
\begin{figure*}[h!]
\centering
\includegraphics[width=120mm]{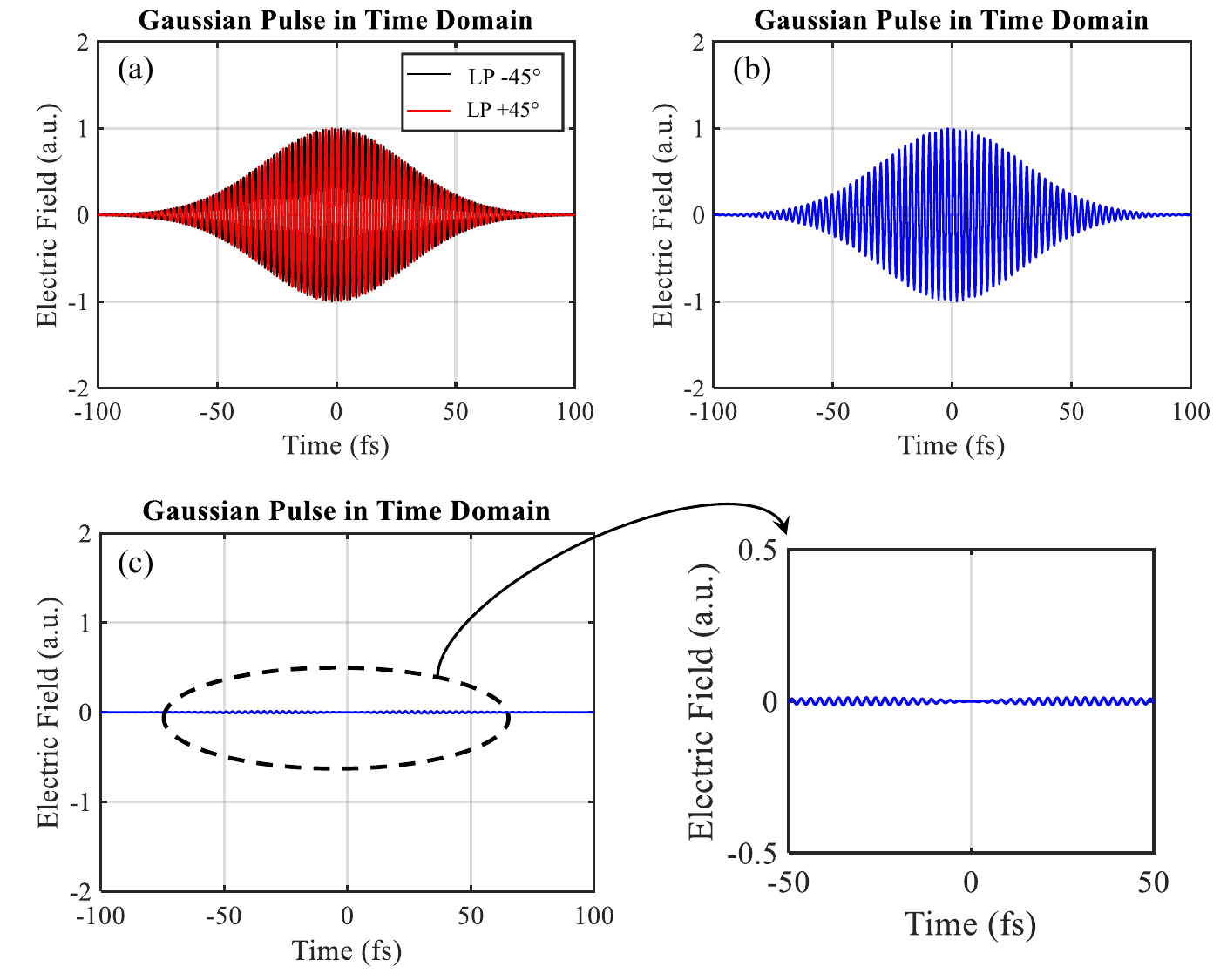}
\caption{ (a) Directly calculated pulse shape after the transmission through a HWP at $45^\circ$. The initial vertical polarization splits into the polarization components at $\pm45^\circ$, marked with a red and a blue color. (b) An amplitude profile of the pulse calculated as the sum of the linear states. 
(c) A residual $y$ polarization state after subtraction of the dominant $x$ polarization. We used real values for the time duration and the central frequency of the pulse. 
 }
\label{pulse_ret}
\end{figure*}
Figure \ref{pulse_ret} visualizes the effect of the weak measurement in time for a short laser pulse. We have used the real values of the laser pulse duration time and the central frequency to calculate the pulse shape after the retardation by a HWP at $\pm 45^\circ$. In Fig.\ref{pulse_ret}a the retarded linear polarization components are shown together (in blue and red color). When summing up their amplitudes a Gaussian pulse exhibits mainly a linear polarization perpendicular to the incident one (see Fig.\ref{pulse_ret}b). Nevertheless, when removing this dominant component, one can recognize a minute signal in the orthogonal state (shown in Fig. \ref{pulse_ret}c. Clearly, this appears to be a feature of time-limited laser pulse. 

\begin{figure*}[ht!]
\centering
\includegraphics[width=120mm]{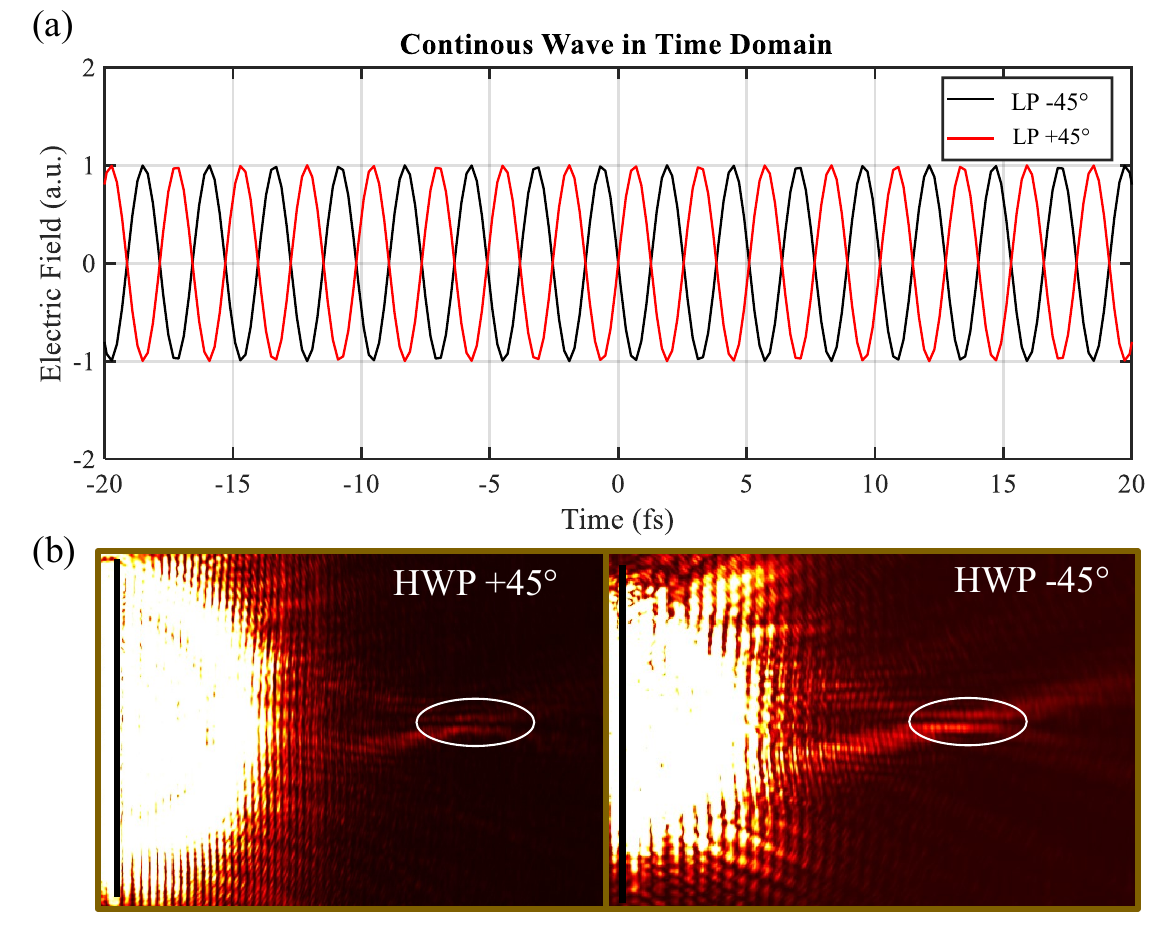}
\caption{ In (a) CW for the LP $+45^\circ$ and $-45^\circ$ are very symmetric, (b) Weak measurement for the HWP $-45^\circ$ and $+45^\circ$ results for the two spots. }
\label{CW}
\end{figure*}

The illustration of the retardation effect on a CW beam is presented in Fig. \ref{CW}a. The $+45^\circ$ and the $-45^\circ$ linear components are retarded from each other by exactly a half of the period which results in a rotated linear polarization. The beam state does not experience any time variation, therefore there is no expected difference between the cases of the HWP rotated at $45^\circ$ and $-45^\circ$. 
We perform a test experiment with a CW laser at the same wavelength of $\lambda_0 = 785\,\mathrm{nm}$ and present the results in Fig.\ref{CW}b. As predicted, no angular or spatial misalignment have been found for the studied cases. 

\section{Measurement scheme for Circular Birefringence and $Q$-plate}
\begin{figure*}[ht!]
\centering
\includegraphics[width=100mm]{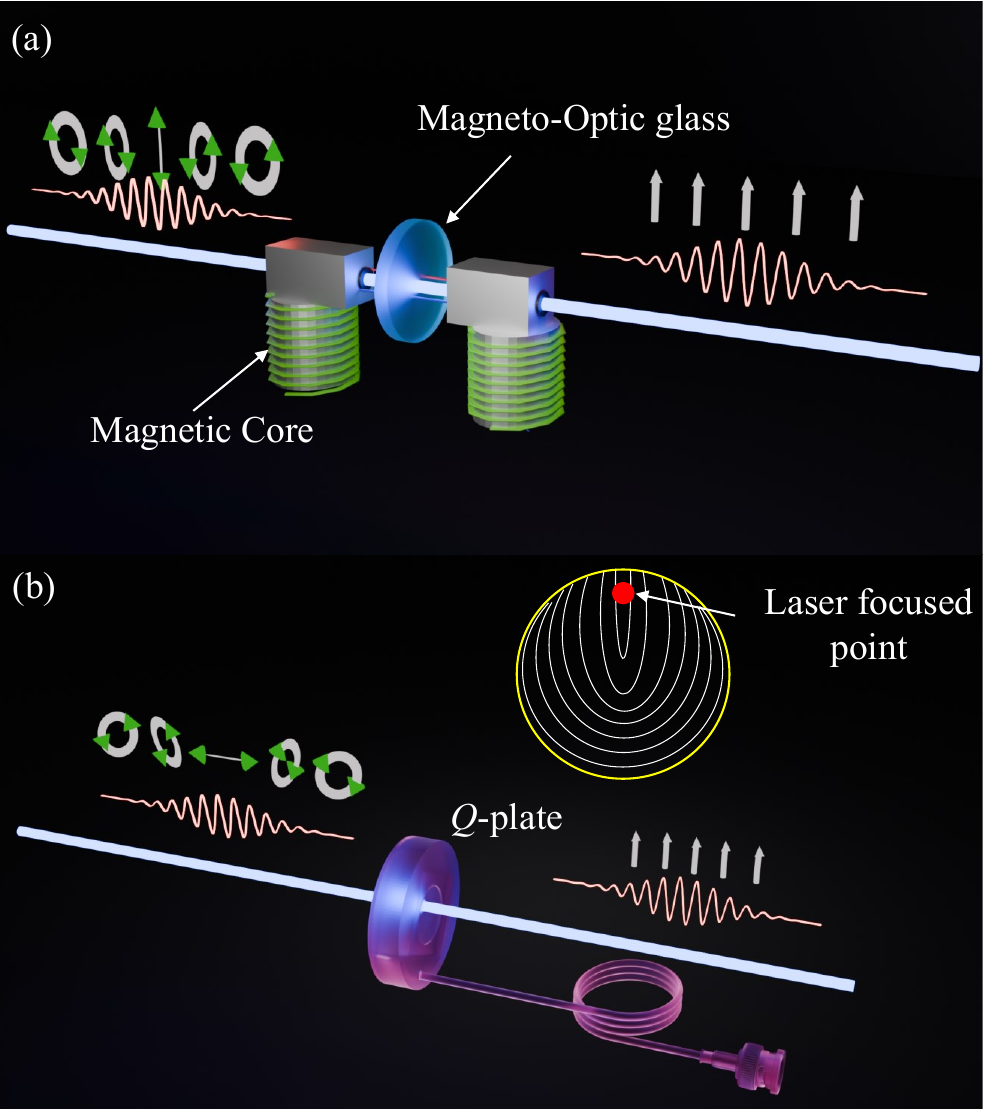}
\caption{ (a) In the schematic setup, when the pulse passes through the MO glass, it generates a circular birefringence, leading to a variation in the susceptibility $\chi$ throughout the pulse. (b) When the pulse then passes through the $Q$-plate, simultaneous rotation of linear polarization and changes in the ellipticity of the polarization are observed.}
\label{SchemeMO}
\end{figure*}
To generate circular birefringence, we used MO glass MR3-2 subjected to a 300 mT external magnetic field, applied using a coil and a funnel that directs the magnetic field along the pulse path (see Fig. \ref{SchemeMO}a). As the pulse passes through the MO glass, it experiences circular birefringence, which introduces ellipticity. To simultaneously generate both linear polarization rotation and ellipticity, we used a $Q$-plate (spiral plate, ARCoptix, Switzerland), as shown in Fig. \ref{SchemeMO}b. This optical element is designed to generate a radial polarization. Accordingly it comprises of a liquid crystal molecules aligned with their primary axis rotating along the azimuthal angle (see inset in Fig. \ref{SchemeMO}b). In the experiment the unexpanded laser beam ($\sim 2\,\mathrm{mm}$ in diameter) passed through a upmost spot in the $Q$-plate (see inset in Fig. \ref{SchemeMO}b). 

\newpage
\bibliography{apssamp}
\end{document}